\documentclass[twocolumn,showpacs,amsmath,amssymb,prb]{revtex4-1}
\usepackage{graphicx}% Include figure files
\usepackage{dcolumn}% Align table columns on decimal point
\usepackage{bm}% bold math
\usepackage[abs]{overpic}
\def\up{\uparrow}
\def\down{\downarrow }
\def\Vec#1{\bm{#1}}
\newcommand{\rmd}{{\rm d}}
\begin{document}

%\preprint{}

\title{ Direct Numerical Demonstration of Sign-preserving Quasiparticle Interference via 
Impurity inside Vortex Core in Unconventional Superconductors }

\author{Yuki Nagai}
\affiliation{CCSE, Japan  Atomic Energy Agency, 6-9-3 Higashi-Ueno, Tokyo 110-0015, Japan}
\affiliation{CREST(JST), 4-1-8 Honcho, Kawaguchi, Saitama, 332-0012, Japan}
\author{Noriyuki Nakai}
\affiliation{CCSE, Japan  Atomic Energy Agency, 6-9-3 Higashi-Ueno, Tokyo 110-0015, Japan}
\affiliation{CREST(JST), 4-1-8 Honcho, Kawaguchi, Saitama, 332-0012, Japan}
\author{Masahiko Machida}
\affiliation{CCSE, Japan  Atomic Energy Agency, 6-9-3 Higashi-Ueno, Tokyo 110-0015, Japan}
\affiliation{CREST(JST), 4-1-8 Honcho, Kawaguchi, Saitama, 332-0012, Japan}

%$^{2}$CREST(JST), 4-1-8 Honcho, Kawaguchi, Saitama, 332-0012, Japan\\
%$^{3}$TRIP(JST), Chiyoda, Tokyo, 102-0075, Japan
%}

\date{\today}% It is always \today, today,
             %  but any date may be explicitly specified

\begin{abstract}

We perform large-scale numerical calculations self-consistently solving the Bogoliubov-de Gennes (BdG) equations 
in the magnetic field together with random impurities to directly demonstrate the typical quasi-particle interference (QPI) 
in the presence of vortices as observed by scanning tunneling microscopy/spectroscopy experiments 
in unconventional superconductors. 
The calculations reveal that vortex itself never works as a scatter causing the QPI pattern but vortex core 
containing impurity brings about the enhancement of the sign-preserving QPI peaks.
Its origin is Andreev bound-states distorted by impurity, and all the measurement findings are consistently 
explained by the scenario based on the numerical results.

%%%%%%%% Check the abstract again after checking all sentences 7/5 11:44 %%%%%%%%%%%%%%%%% 
\end{abstract}

% PACS, the Physics and Astronomy
                             % Classification Scheme.
%\keywords{Suggested keywords}%Use showkeys class option if keyword
                              %display desired
\maketitle
%%%%%%%%%
\section{Introduction}
Since the discovery of High-Tc cuprate superconductors, the pairing mechanism
has been a central issue in condensed matter physics over two decades.
The superconducting gap symmetry was 
one of the most important clues to identify the mechanism.
Historically, the symmetry in High-Tc cuprate superconductors 
was proven to be $d$-wave by a direct observation 
of the half-fluxon in the tri-crystal junction and supported by other measurements\cite{Tsuei}.
Recently, a new type of symmetry, i.e., the sign-reversing $s$-wave pairing symmetry 
has been proposed as a candidate in brand-new iron-based High-Tc superconductors. 

The measurement of quasi-particle interference (QPI) patterns 
by the scanning tunneling microscopy/spectroscopy (STM/STS)
is now listed as a powerful tool directly probing the pairing symmetry 
\cite{Hoffman,McElroy,Hanaguri,Kohsaka,HanaguriSci,Pereg-Barnea08,Maltseva}. 
The QPI pattern is obtained by Fourier-transforming the ratio $Z({\bm r},V) = g({\bm r},V)/g({\bm r},-V)$ of 
the conductance maps $g({\bm r},V) = dI/dV({\bm r},V)$ for the position ${\bm r}$ and the bias voltage $\pm V$. 
So far, it has been widely accepted that a relative sign difference inside superconducting order parameter can 
be directly 
detected by the magnetic field dependence of its Fourier transformation $|Z({\bm q},\omega)|$.
The reason is that if a vortex works as a magnetic impurity then 
the number of magnetic impurities increases with increasing the magnetic field and 
the sign-preserving scattering \cite{Hanaguri} dominates over the sign-reversing one. 
However, it is just an intuitive idea, and there is no 
direct theoretical confirmation except for an indirect support 
through a perturbative approach\cite{Maltseva}. 
%%%%%%%%%%%%%%% 2011 7/5 12:06 %%%%%%%%%%%%%%%%%%%%%%%%%%%%%%%%%%%%%%%%%%%%
Generally, vortex breaks the translational symmetry and makes  
low-energy Andreev bound states inside the core.
Thus, the vortex is never a simple impurity and beyond the reach of any perturbative approaches.
In iron-based superconductors, the QPI experimental result \cite{HanaguriFe} is 
presently regarded as one of a few crucial  
facts supporting the sign-reversing $s$-wave symmetry.
We have to rush to confirm the intuitive assumption. 

%%%%%%%%%%%%%%%%%% 2011 7/9 18:56 %%%%%%%%%%%%%%%%%%%%%%%%%%%%%%%%%%%%%%%%%%

In this paper, we therefore investigate QPI in the presence of vortices through 
a direct numerical calculation on the self-consistent BdG equations.
A main finding of the present paper is that vortex works as a scatter breaking the time-reversal symmetry 
only when impurity is captured by the core.
Thus, this finding indicates that vortex pinning, i.e., so-called the core pinning is 
crucial 
in supporting 
%for consistency with 
the previous QPI scenario in the magnetic field.

%%%%%%%%%%% 3/9 2:28 %%%%%%%%%%%%%%%%%%%%%%%%%%%%%%%%%%%%%%%%%%%%%%%% 

To numerically obtain the QPI map, we need to perform a large-scale calculation in real-space, since the QPI map 
is obtained by Fourier transforming a sufficiently wide real-space data. 
However, the computer resource to numerically solve the BdG equations 
becomes too huge, as long as one employs the conventional Hamiltonian-matrix diagonalization-scheme whose
computational cost increases in $N^3$ manner, where $N$ is the number of grid points in real-space.
In this paper, we instead implement 
the Chebyshev-polynomial expansion scheme\cite{Covaci,Tanaka,Kunishima} 
as a self-consistent solver of the BdG equations\cite{NagaiOta}.
We stress that the present real-space area is wider than 10-nm scale square, which 
is much beyond the conventional size.

\section{Model and Method}

The target tight-binding model Hamiltonian %for the direct demonstration is written as 
${\cal H} = {\cal H}_{\rm BCS} + {\cal H}_{\rm imp}$. 
%\begin{align}
%{\cal H} = {\cal H}_{\rm BCS} + {\cal H}_{\rm imp}. 
% {\cal H}_{\rm BCS} &= - \sum_{ij,\sigma} (\tilde{t}_{ij} + (\mu + V^{\rm imp}_{i} ) \delta_{ij} )c^{\dagger}_{i \sigma}c_{j \sigma} 
%+ \sum_{ij} \left[ \Delta_{ij} c^{\dagger}_{i \up} c_{j \down}^{\dagger} + {\rm h.c.}\right].
%\end{align}
Here, BCS Hamiltonian, ${\cal H}_{\rm BCS} = - \sum_{ij,\sigma} (\tilde{t}_{ij} + \mu  \delta_{ij} )c^{\dagger}_{i \sigma}c_{j \sigma} 
+ \sum_{ij} \left[ \Delta_{ij} c^{\dagger}_{i \up} c_{j \down}^{\dagger} + {\rm h.c.}\right]$, where 
$c^{\dagger}_{i\sigma}$ creates an electron with spin $\sigma$ at site $i$ and $\mu$ denotes the chemical potential. 
The hopping integrals $\tilde{t}_{ij}$ include the Peierls phase factor 
$\tilde{t}_{ij} = t_{ij} \exp \left[ i \frac{\pi}{\phi_{0}} \int_{{\bm r}_{i}}^{{\bm r}_{j}} {\bm A}({\bm r}) \cdot d {\bm r} \right]$ 
in the presence of the magnetic field, 
where ${\bm A}({\bm r})$ is the vector potential and $\phi_{0} = hc/2e$ is the flux quantum. 
The impurity part of the Hamiltonian can be written as ${\cal H}_{\rm imp} = 
\sum_{i,\sigma} V_{i}^{\rm imp} c^{\dagger}_{i \sigma}c_{i \sigma}$. 
One diagonalizes  
%the above quadratic  Hamiltonian 
${\cal H}$ 
to solve the BdG equations written as 
\begin{align}
\sum_{j} 
\left(\begin{array}{cc}\hat{K}_{i,j} & \hat{\Delta}_{i,j} \\
\hat{\Delta}_{i,j}^{\dagger} & -\hat{K}_{i,j}^{\ast}\end{array}\right) 
\left(\begin{array}{c}u_{\alpha}({\bm r}_{j}) \\
v_{\alpha}({\bm r}_{j})
\end{array}\right) 
= E_{\alpha} 
\left(\begin{array}{c}u_{\alpha}({\bm r}_{i}) \\
v_{\alpha}({\bm r}_{i})
\end{array}\right). \label{eq:bdgeq}
\end{align}
Here,  
$K_{i,j} = - \tilde{t}_{ij} - (\mu - V^{\rm imp}_{i})  \delta_{ij}$, and 
$\Delta_{i,j}=\Delta({\bm r}_{i},{\bm r}_{j}) = V_{ij} \sum_{\alpha}^{2 N} u_{\alpha}({\bm r}_{j}) 
v_{\alpha}^{\ast}({\bm r}_{i}) f(E_{\alpha})$,
where $N$ is the number of the lattice sites, $V_{ij}$ denotes the pairing interaction, and $f(x)$ is 
the Fermi distribution function.
%\end{align}
%with 
In this paper, the hopping is restricted only in the nearest neighbor ($t_{ij} = t$) for simplicity. 
The chemical potential $\mu = -1.5t$, and the pairing interaction is given only on the link of the 
nearest neighbor, $V_{ij} = V = -2.2t$. 
We self-consistently calculate $d_{x^{2}-y^{2}}$-wave order parameter, $\Delta_{d}({\bm r}_{i}) = V(\Delta_{\hat{x},i} 
+ \Delta_{-\hat{x},i} - \Delta_{\hat{y},i} - \Delta_{-\hat{y},i})/4$ 
with $\Delta_{\pm \hat{e},i} = \Delta(\Vec{r}_{i},\Vec{r}_{i}\pm\Vec{\hat{e}})\exp \left[ i \frac{\pi}{\phi_{0}} \int^{({\bm r}_{i}
+{\bm r}_{i} \pm \Vec{\hat{e}})/2}_{{\bm r}_{i}} {\bm A}({\bm r}) \cdot d {\bm r} \right]$\cite{Takigawa}, where 
$\hat{x}$ and $\hat{y}$ denote the unit vectors in a square lattice. 
%$\Delta({\bm r}_{i},{\bm r}_{j})$ is the bond-dependent superconducting gap, which should be self-consistently calculated. 

Let us briefly show how to solve self-consistently the BdG equations (\ref{eq:bdgeq}) including the gap equation with use of 
%spectrum-density polynomial method\cite{NagaiOta}. 
the Chebyshev-polynomial expansion scheme. 
The mean-field can be expressed as 
$\langle c_{i} c_{j} \rangle = 
- \frac{1}{2 \pi i} \int_{- \infty}^{\infty} \rmd \omega f(\omega) 
\vec{e}(j)^{\rm T} \hat{d}(\omega) \vec{h}(i)$, where $[\vec{e}(i)]_{\gamma} = \delta_{i,\gamma}$, and 
$ [\vec{h}(i)]_{\gamma} = \delta_{i+N,\gamma}$. 
The spectral density $\hat{d}(\omega)$ is given by $\hat{G}^{\rm R}(\omega)- \hat{G}^{\rm A}(\omega)$ whose
Dirac's delta functions are expanded by a series of Chebyshev-polynomials\cite{NagaiOta}. 
Then, one can rewrite the gap equation at zero-temperature as 
\begin{align}
\Delta({\bm r}_{i},{\bm r}_{j}) &= -\frac{2 V_{ij}}{\pi}
\sum_{n = 1}^{n_{c}}
\vec{e}(j)^{\rm T} \vec{h}_{n}(i) 
\frac{\sin[n \arccos(-b/a)]}{n}
, \label{eq:cct} 
\end{align}
where $n$ is the order of 
the Chebyshev-polynomial, and $n_{c}$ denotes a cutoff parameter. 
$\vec{h}_{n}(i)$ is calculated by the recurrence formula,  
$\vec{h}_{n+1}(i) = 2 ({\cal H} - \hat{1} b)/a \vec{h}_{n}(i) - \vec{h}_{n-1}(i)$ and $\vec{h}_{0}(i) = \vec{h}(i)$ 
and $\vec{h}_{1}(i) = 2 ({\cal H} - \hat{1} b)/a \vec{h}(i)$, where 
the renormalized factors $a$ and $b$ are set in the order of the band-width, e.g., $b = - \mu$, respectively. 
We point out that calculation results are insensitive to choice of these parameters \cite{Tanaka,NagaiOta}.
This drastically reduces the self-consistent calculation costs 
since the calculation algorithm is 
%due to 
perfectly free from the use of any diagonalization scheme.
%%%%%%%%%%%%%%%%%%%%%%%%%%%%% 7/09 23:04 %%%%%%%%%%%%%%%%%%%%%%%%%%%%%%%%%%%%%%%%%%% 
We always set $n_{c} = 1000$ through the present calculations, 
which is confirmed to be enough\cite{NagaiOta}. 
Since each grid point is completely independent in the vector formula, 
we can efficiently solve these equations by using parallel multi-core computers.

The calculation target is $160 \times 160$ square lattice, which conventionally requires 
a full diagonalization of $51200 \times 51200$ matrix. % when using the conventional diagonalization scheme. 
If the lattice constant $\sim \AA$, then its real-scale area is beyond 10 nm square.  
In this case, it is a quite hard task for the conventional diagonalization scheme
to self-consistently solve the problem, while the present 
scheme can finish the self-consistent calculation during about 10 minutes when using 128 cores. 
%%%%%%%%%%%%%%%%%%%%% 7/9 1:18 %%%%%%%%%%%%%%%%%%%%%%%%%%%%%%%%%%%%

\section{Results}

Here, let us show how to demonstrate that QPI in the presence of the magnetic field 
actually detects the sign-preserving quasi-particle scattering. 
First, we examine the quasi-particle scattering 
by non-magnetic impurities without vortices. 
Second, we study QPI for a vortex lattice without any impurities to 
know whether vortex works as a magnetic impurity. 
Third and fourth, we investigate two vortex systems with random impurities.
In the third case, all impurities are away from any vortex cores.
On the other hand, one impurity locates inside the vortex core in the forth case. %
Through the comparison between the third and fourth cases, 
we find that impurity locating inside 
the vortex core is essential to identify the relative phase difference of the
superconducting order parameter.

\begin{figure}[tb]
  \begin{center}
    \begin{tabular}{p{\columnwidth}}%  p{28mm}}
      \resizebox{0.5 \columnwidth}{!}{\includegraphics{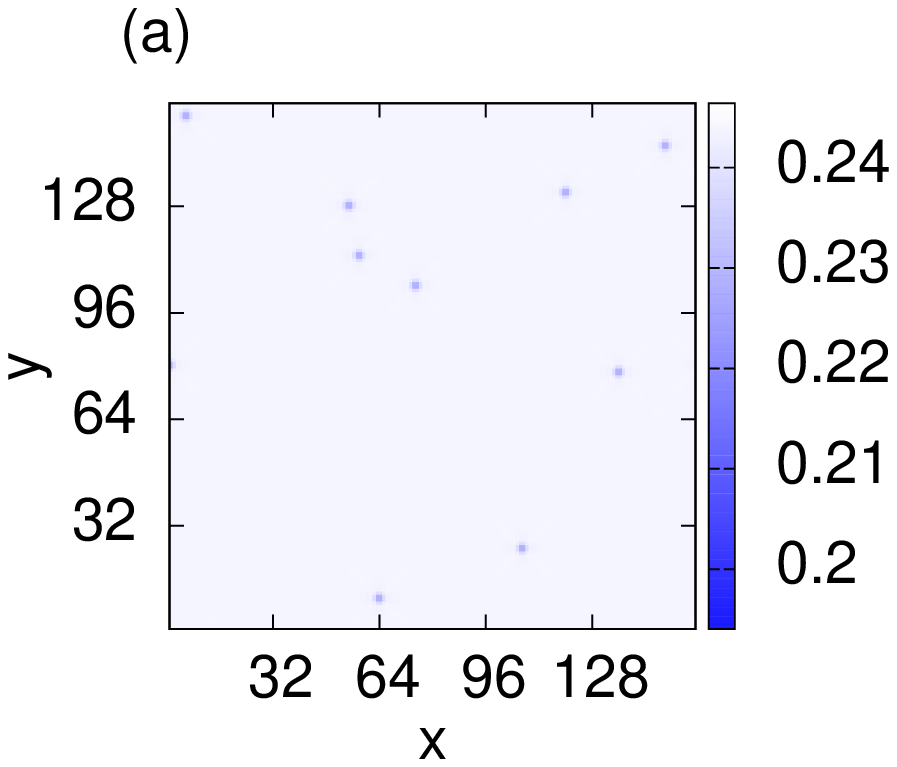}} %\\
      \resizebox{0.45  \columnwidth}{!}{\includegraphics{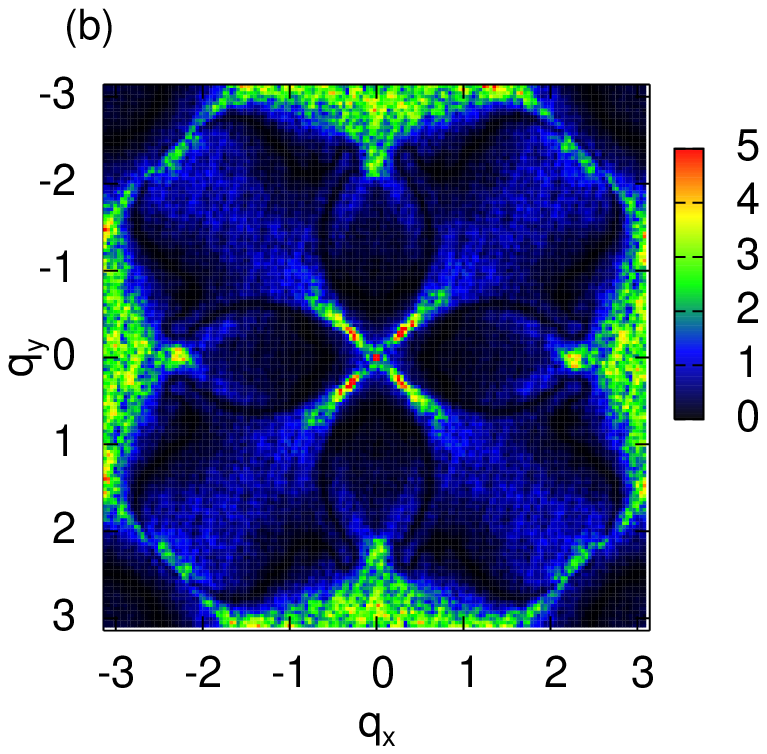}} % &
   %   \resizebox{30mm}{!}{\includegraphics{Qmap100930_qzpi.eps}} 
    \end{tabular}
\caption{\label{fig:fig1}
(Color online) 
(a): A spatial profile of the $d_{x^{2}-y^{2}}$-wave order-parameter amplitude $|\Delta_{d}({\bm r}_{i})|$.
(b): A $\Vec{q}$-space profile of $Z({\bm q},E)$, which is 
obtained by Fourier transforming the quantity $Z({\bm r},\omega) \equiv N({\bm r},\omega)/N({\bm r},-\omega)$ at the energy $\omega = 0.04t$. 
The system size is  $160 \times 160$ square grids 
with 10 randomly distributed impurities without vortex. }
  \end{center}
\end{figure}

%%%%%%%%%%
%%%%%%%%%%%%%%%%% 7/10 12:00 %%%%%%%%%%%%%%%%%%%%%%%%%%%%%%%%%%%%%
%
At first, let us examine the QPI with no vortex.  
We introduce randomly-distributed impurities as $V^{\rm imp}_{i} = 0.3 t \delta({\bm r} - {\bm r}^{\rm imp,i})$  ($i = 1, \cdots 10$). 
As shown in Fig.~\ref{fig:fig1}(a), the order parameter amplitude $|\Delta_d ({\bf r}_{i})|$ shows 
relatively small values only around the impurity sites. 
In this case, as seen in Fig.\ref{fig:fig1}(b), the quantity $Z({\bm q},E)$, i.e., the Fourier transformation of 
$Z({\bm r},\omega) ( \equiv N({\bm r},\omega)/N({\bm r},-\omega) )$ is equivalent to those 
in the previous theoretical studies using the diagrammatic approach \cite{Maltseva}.
This map is regarded as a typical QPI pattern, which clearly reflects the scattering feature of 
quasi-particles on the Fermi surface (see the left panel of Fig.4 for the present Fermi surface).

%%%%%%%%

\begin{figure}[t]
  \begin{center}
    \begin{tabular}{p{\columnwidth}}%  p{28mm}}
      \resizebox{0.5 \columnwidth}{!}{\includegraphics{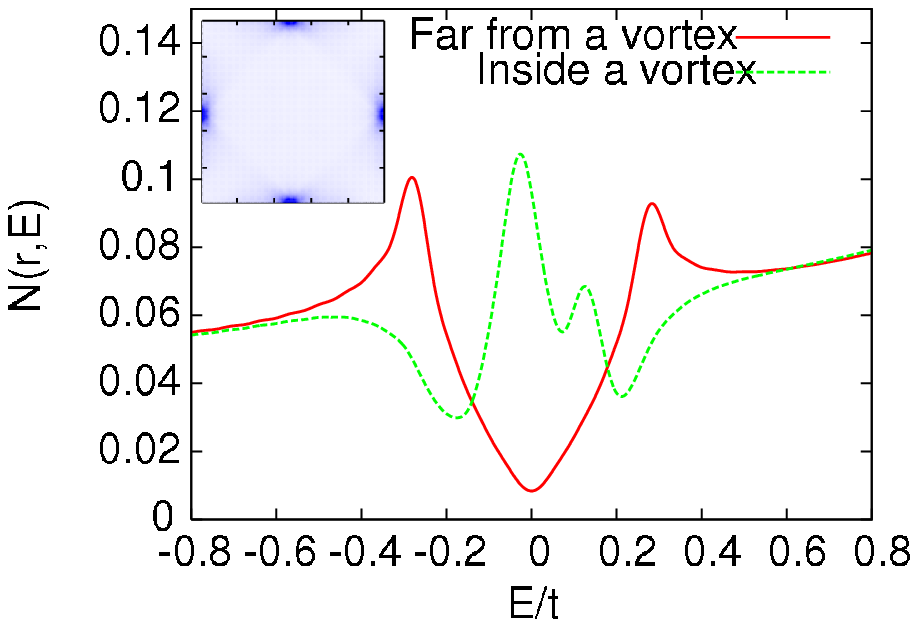}} %\\
      \resizebox{0.5  \columnwidth}{!}{\includegraphics{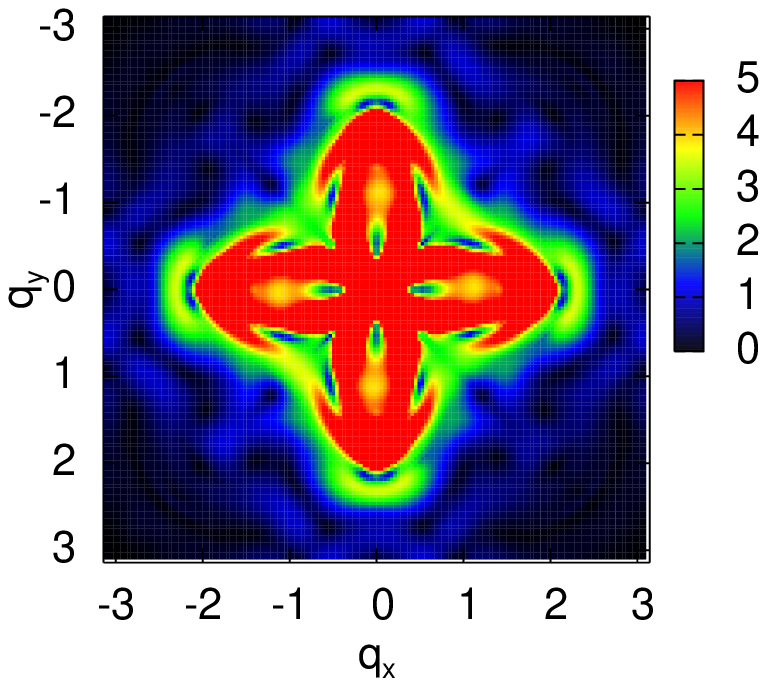}} % &
   %   \resizebox{30mm}{!}{\includegraphics{Qmap100930_qzpi.eps}} 
    \end{tabular}
\caption{\label{fig:fig2}
(Color online) 
Left panel: Local densities of states far from a vortex core and inside a vortex core without any impurities. 
Inset: A spatial profile of the $d_{x^{2}-y^{2}}$-wave order-parameter amplitude $|\Delta_{d}({\bm r}_{i})|$.
Right panel: A $\Vec{q}$-space profile of $Z({\bm q},E)$. 
For the definition $Z({\bm q},E)$ and the system size, see the caption of Fig.~\ref{fig:fig1}. 
%Inset: Absolute values of the $d_{x^{2}-y^{2}}$-wave order parameter $|\Delta_{d}({\bm r}_{i})|$.
}
  \end{center}
\end{figure}

Second, we study a vortex lattice without any impurities. 
According to the standard way introducing vortices\cite{Wang, Takigawa}, 
we make a square vortex lattice, whose unit vector is in the 
direction of 45$^{\circ}$ from $a$-axis of 
the original tight-binding model (see the inset of Fig.~\ref{fig:fig2}). 
Then, the local density of states (LDOS), $N(r,E)$ is calculated and the gap edge is 
found to be $\sim 0.25t$ as seen in the left panel of Fig.~\ref{fig:fig2}. 
The QPI map is displayed in the right panel of Fig.~\ref{fig:fig2}, where we note that 
the map is irrelevant to QPI. 
Namely, one can not find out any characteristic wave-vector peaks arising from the impurity 
scattering in contrast to Fig.~\ref{fig:fig1}. In the case, %the vortex lattice without any disorders, 
Andreev bound states inside the vortex core just produce such a pattern only around $q \sim 0$. 
Thus, it is found that vortex can not be regarded as a magnetic impurity scatter. 
%%%%%%%%%%%%% 7/10 12:32 %%%%%%%%%%%%%%%%%%%%%%%%%%%%%%%%%%%%%%%%%%%%%%

%$\Vec{q}$-map of $Z({\bm q},E)$ 
%This $\Vec{q}$-map is originated from the localized states near a vortex core. 
%Here, it should be noted that this map on the clean vortices is not almost relevant to QPI. 

%%%%%%%%

\begin{figure}[t]
  \begin{center}
    \begin{tabular}{p{\columnwidth}}%  p{28mm}}
%      \resizebox{0.5 \columnwidth}{!}{\includegraphics{Fig3a.eps}} %\\
      %\begin{overpic}
      %\end{overpic}
      \resizebox{0.5 \columnwidth}{!} 
      {
\includegraphics{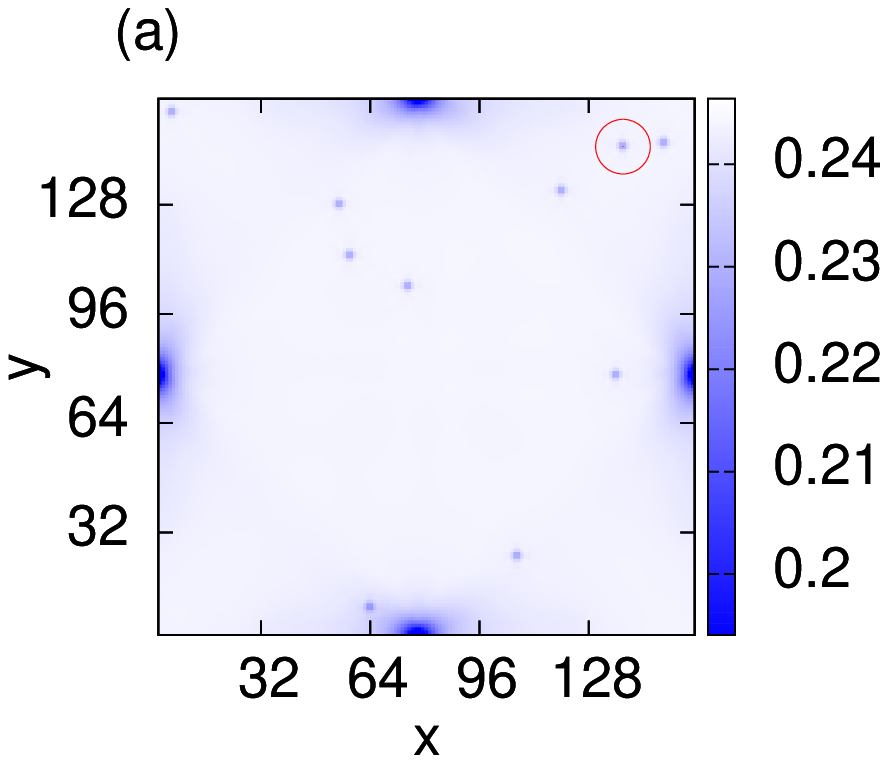}
      } %\\
       \resizebox{0.45  \columnwidth}{!}{\includegraphics{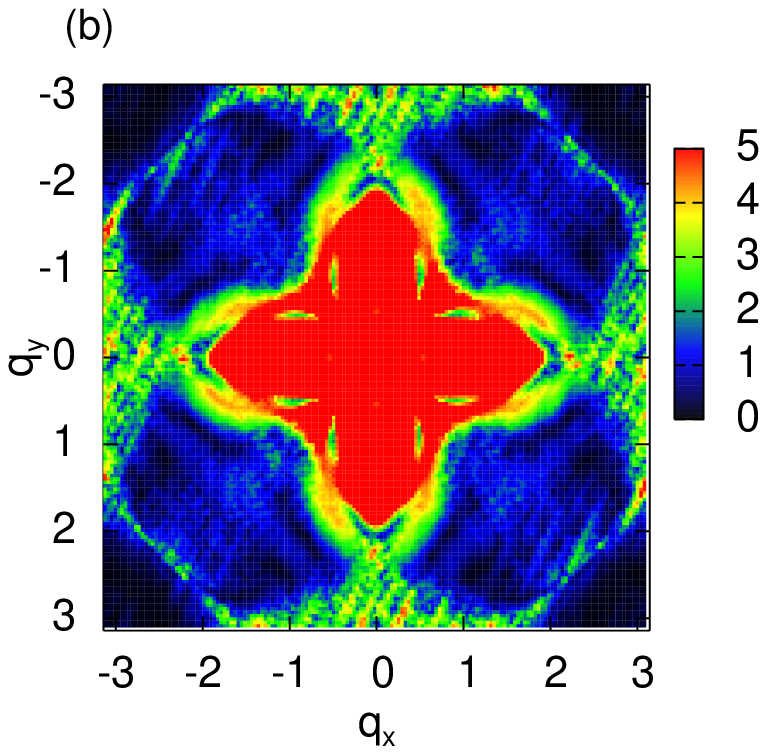}}  \\
   %   \resizebox{30mm}{!}{\includegraphics{Qmap100930_qzpi.eps}} 
         \resizebox{0.5 \columnwidth}{!}{\includegraphics{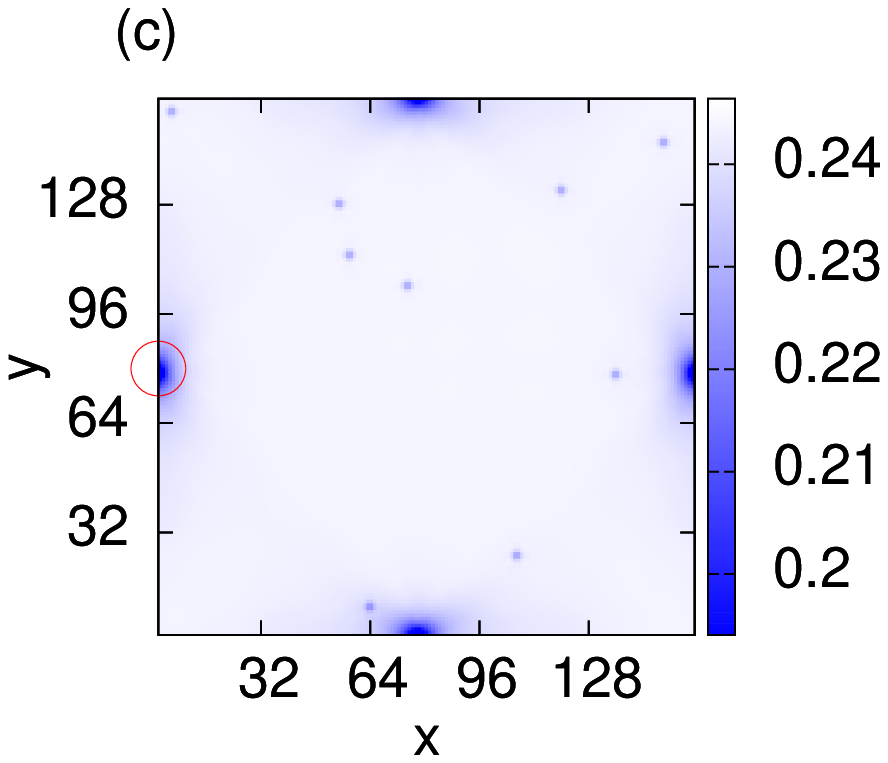}} %\\
       \resizebox{0.45  \columnwidth}{!}{\includegraphics{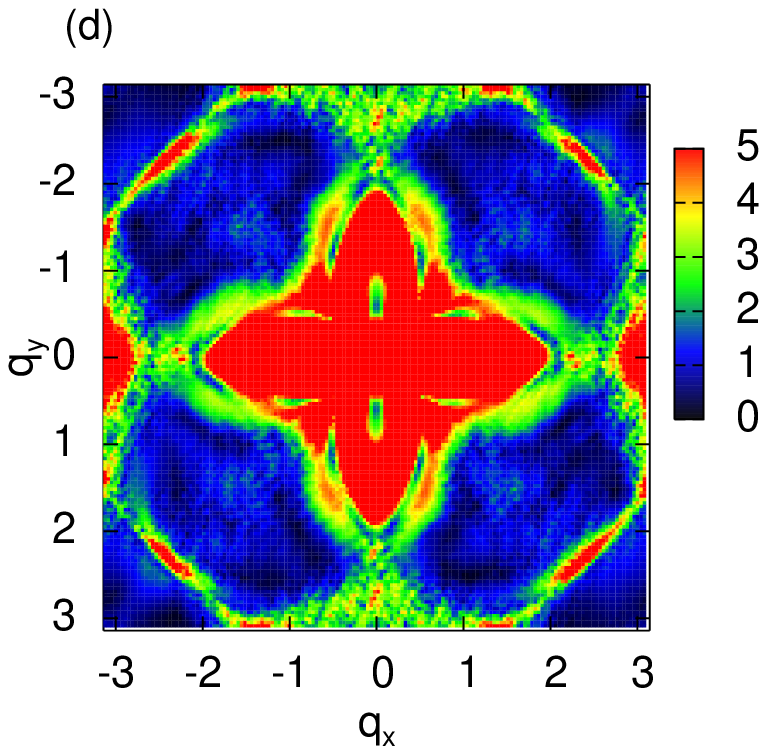}}  
%      \resizebox{30mm}{!}{\includegraphics{Qmap100930_qzpi.eps}} 
    \end{tabular}
\caption{\label{fig:fig3}
(Color online) 
(a) and (c): 
Spatial profiles of the $d_{x^{2}-y^{2}}$-wave order-parameter amplitude $|\Delta_{d}({\bm r}_{i})|$.
%Absolute values of the $d_{x^{2}-y^{2}}$-wave order parameter $|\Delta_{d}({\bm r}_{i})|$.
(b) and (d): $\Vec{q}$-space profiles of $Z({\bm q},E)$. 
For the definition $Z({\bm q},E)$, see the caption of Fig.~\ref{fig:fig1}. 
We consider the $160 \times 160$ square lattice system with 10 randomly distributed impurities with vortices. 
(a) and (b):  Any impurities are not located inside vortices (``unpinned vortex'').  
(c) and (d):  One impurity is located inside a vortex (``pinned vortex'). 
}
  \end{center}
\end{figure}

Now, let us introduce random impurities into the vortex lattice. 
As the third case, 10 impurities are randomly distributed, but 
all impurities are away from any vortex cores.
In this case, their impurity potential is not so deep and 
their density is so dilute that the vortex lattice is not almost distorted compared 
to the case without impurities. 
Such a situation is called "floating lattice", in which an
energy gain by the vortex lattice formation dominates over 
that by the vortex core pinning. 
As shown in Fig.~\ref{fig:fig3}(b), the QPI pattern in the floating lattice 
is almost regarded as a sum of QPI's in the first and second cases.

\begin{figure}[t]
  \begin{center}
    \begin{tabular}{p{\columnwidth}}%  p{28mm}}
      \resizebox{0.45 \columnwidth}{!}{\includegraphics{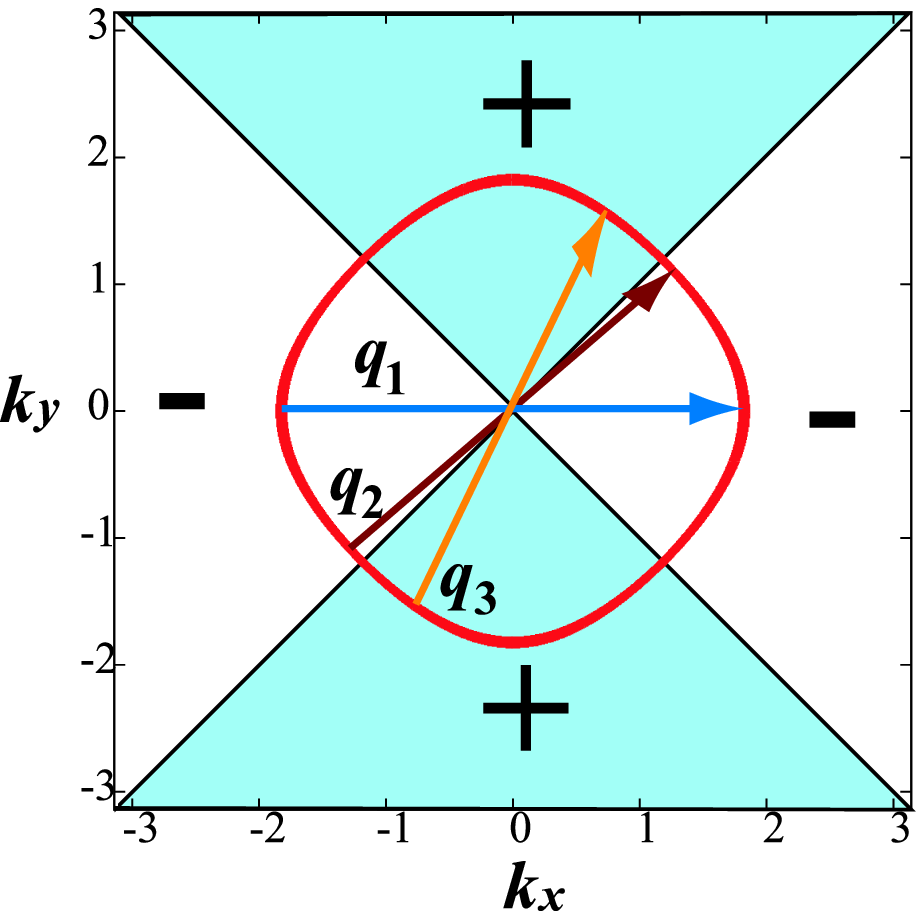}} %\\
      \resizebox{0.50  \columnwidth}{!}{\includegraphics{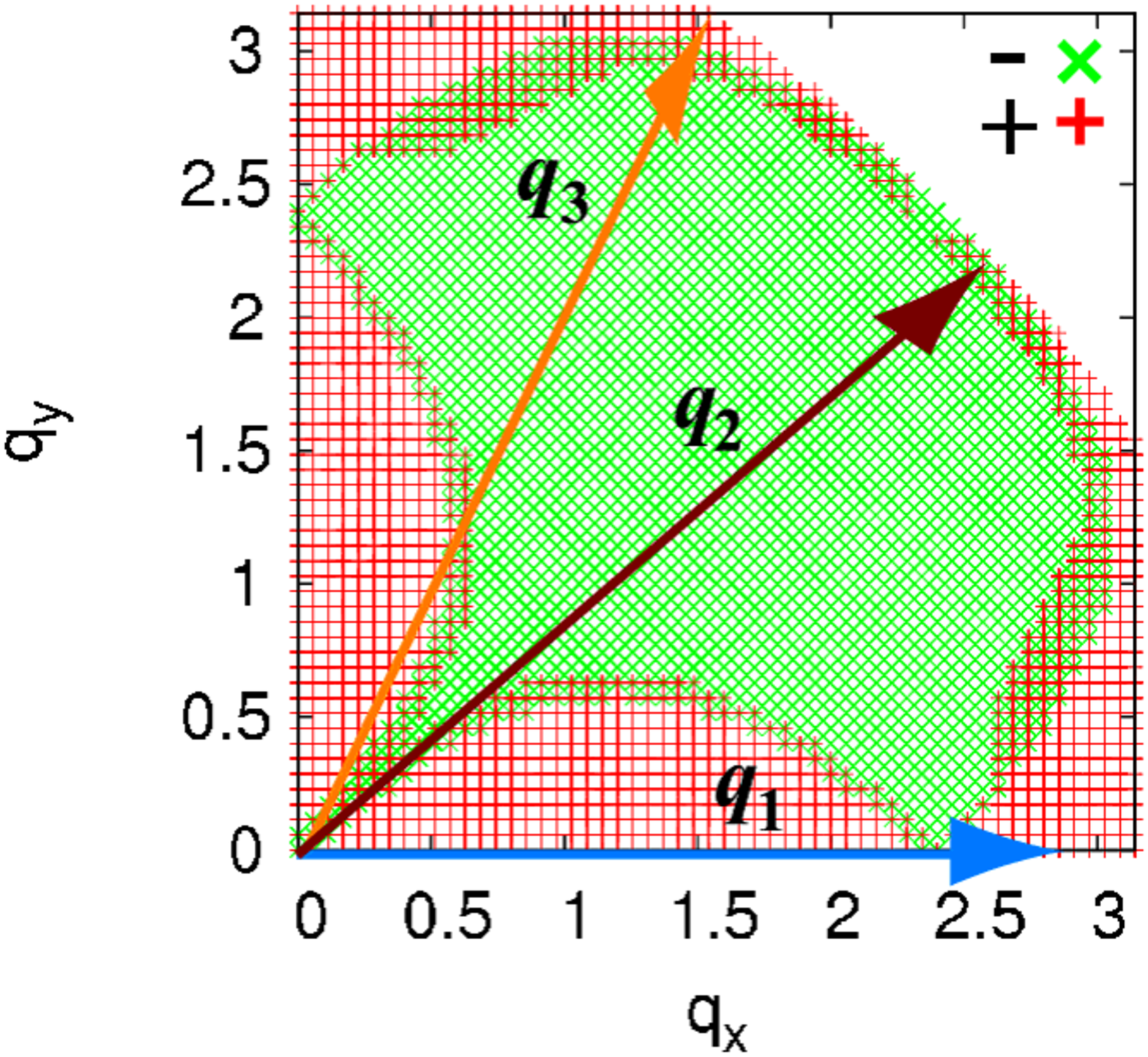}} % &
   %   \resizebox{30mm}{!}{\includegraphics{Qmap100930_qzpi.eps}} 
    \end{tabular}
\caption{\label{fig:fig4}
(Color online) 
Left panel: 
A $\Vec{k}$-space profile of the Fermi surface (red solid curve) in the square lattice model with $\mu = -1.5t$ 
and the white and shaded areas represent signs 
of $d$-wave superconducting 
gap on $\Vec{k}$-space. 
The arrows denote 
typical sign-preserving scattering $\Vec{q}$-vectors (See the text). 
Right panel: 
A $\Vec{q}$-space profile of the sign preserving (+) and reversing (-) on the superconducting gap.
%Distribution of the signs of scatterings in $\Vec{q}$-space.
}
  \end{center}
\end{figure}

Here, let us move each impurity site into the vortex core in order to 
create the so-called "pinned lattice". 
In this case, we find that the calculated Free energy is 
reduced compared to that of floating lattice. 
As the fourth case, we move an impurity marked by the red line circle in Fig.~\ref{fig:fig3}(a) 
to the marked vortex core as shown in Fig.~\ref{fig:fig3}(c). 
Then, the marked vortex of Fig.~\ref{fig:fig3}(c) is assigned to "pinned vortex", and 
the QPI in the presence of the pinned vortex is displayed in Fig.~\ref{fig:fig3}(d). 
%We confirm that the location inside the vortex core does not change our main results. 
%%%%%%%%%%%%%%%% 7/10 15:11 %%%%%%%%%%%%%%%%%%%%%%%%%%%%%%%%%%%%%%%%%%%%%%%%
By comparing Fig.~\ref{fig:fig3}(d) with Fig.~\ref{fig:fig3}(b), 
one finds that intensified points are located close to ${\bm q}_{1} = (\pi,0)$, ${\bm q}_{2} \sim (2.3,2.3)$ 
and ${\bm q}_{3} \sim (1.2,\pi)$.
As these ${\bm q}$'s are displayed in the left panel of Fig.~\ref{fig:fig4},  
${\bm q}_{1}$, ${\bm q}_{2}$ and ${\bm q}_{3}$, correspond to 
the scattering vectors represented by ${\bm k}_{F}^{f}(0) - {\bm k}_{F}^{i}(\pi)$, 
${\bm k}_{F}^{f}(\pi/4) - {\bm k}_{F}^{i}(5 \pi/4)$, and ${\bm k}_{F}^{f}(3 \pi/8) - {\bm k}_{F}^{i}(11\pi/8)$, respectively, where
${\bm k}_{F}^{f}(\theta)$ and ${\bm k}_{F}^{i}(\theta)$ are 
the Fermi wave vectors after and before the scattering, and $\theta$ denotes the position angle on the Fermi surface. 
One can finds that ${\bm q}_{1}$ and ${\bm q}_{3}$ vectors mean sign-preserving scatterings. 
On the other hand, one notices for $\Vec{q}_{2}$-vector that one needs a numerical check in judging  
whether $\Vec{q}_{2}$ really corresponds to a sign-preserving scattering or not.
The result is displayed in the right panel of Fig.~\ref{fig:fig4}, which distinguishes the sign-preserving 
with the reversing one in $\Vec{q}$-space quadrant by actually examining $\Vec{q}$-vectors around
the Fermi surface ($\pm 0.1t$). 
From the right panel of Fig.~\ref{fig:fig4}, it is found that $\Vec{q}_{2}$ is also a
sign-preserving vector.
%As shown in Fig.~\ref{fig:fig4}, these all vectors are the sign-preserving scattering vectors. 
These facts indicate that the sign-preserving scatterings are intensified by the emergence of the pinned vortex core. 
%-------110928 added
In addition,
we should note that the vortex arrangement influenced by the vortex pinning
is not important for QPI phenomena,
since these origins are localized near a vortex-core and an impurity. 
We also note that these QPI phenomena 
are not originated from the disordered effects since we put only ten impurities 
whose density is 0.04\% in our system, which can be assigned as a dilute-impurity doped case.
%--------------------------
%In addition, we should note that the vortex arrangement is not important, since these results are originated from the local phenomena between a vortex and an impurity. 
In reality, the number of pinned vortex cores increases when increasing the magnetic field.
Then, one can expect that only the sign-preserving 
$\Vec{q}$-points grow in the QPI pattern.   
We emphasize that it is consistent with the measurement results.
The presence of pinned vortices is essential for the phase sensitive detection of 
the superconducting order parameter. 
Indeed, such sign-preserving scattering peaks are non-observable in too clean sample\cite{HanaguriPri}.

%%%%%%%%

%%%%%%%%

%%%%%%%%%%%% 7/10 19:58 %%%%%%%%%%%%%%%%%%%%%%%%%
\section{Discussion}

Finally, let us discuss the present numerical scheme and the obtained QPI results. 
The first issue is an advantage of the  
Chebyshev-polynomial expansion scheme in self-consistently calculating the BdG equations. 
In order to obtain a QPI map by Fourier transforming a real-space conductance map, 
one needs to solve the BdG equations on a sufficiently large real-space. 
Thus, there has been so far no study to directly calculate QPI's. 
This clearly indicates that the conventional diagonalization scheme is never 
a practical way in studies like the present topic and an alternative 
more efficient scheme is demanded. 
Very recently, Covaci {\it et al.}, have proposed 
a quite efficient scheme based on the Chebyshev-polynomial expansion
to examine inhomogeneous superconductivity \cite{Covaci}.
The main idea is an expansion of the Green's function by a set of Chebyshev-polynomials, which 
guarantees a drastic reduction of the calculation cost. 
However, Covaci {\it et al.}, left a self-consistent calculation as 
the future work. Thus, the present work is the first self-consistent 
large-scale BdG calculation including 
the vector potential.  
We confirm in the present case that the computational cost totally scales with ${\cal O}(N^{2})$ 
on the real-space grid size $N$ and find that 
the number of the self-consistent iterations 
is almost equivalent with that in the matrix diagonalization. 
This means that the calculation cost is reduced to $\sim 1/N$ compared to the 
diagonalization scheme.  
Moreover, we developed several techniques in order to 
accelerate the calculation (see Ref.\cite{NagaiOta} for the details).
The second issue is an origin of the present QPI pattern in the presence of the pinned vortex. 
Previously, the resonant Andreev scattering 
has been regarded as the scattering mechanism around a vortex \cite{Maltseva,Nunner}.
However, our numerical calculations reveal that 
the spatial variation of the order parameter induced by a vortex 
does not work as an effective scatter. 
In fact, as seen in Fig.~\ref{fig:fig2}, 
we can not find out any characteristic wave-vector peaks unlike the typical QPI map as   
Fig.~\ref{fig:fig1}. 
Thus, the QPI origin of the pinned lattice is explained as follows. 
The Andreev bound states around a vortex core
are slightly distorted by impurity inside the vortex core\cite{note}.
%%%%%%%%%%%%% 7/10 20:30 %%%%%%%%%%%%%%%%%%%%%%%%%%%%%%%%%%%%%%%%%%
A set of such distorted states 
scatter quasi-particles like an impurity while still keeps 
the angular momentum breaking the time reversal symmetry. 
Thus, we easily notice that 
a pinned vortex becomes a sign-preserving scatter and the number of such scatters 
increases with increasing 
the magnetic field. 
On the other hand, the number of the sign non-preserving 
scatter contrarily decreases with increasing the magnetic field, since the number of 
the scalar impurities outside the vortex core decreases with increasing the magnetic field. 
These consequences are consistent with the experimental results. 
In addition, our discussion can be easily applicable to the iron-based superconductors\cite{HanaguriFe}, since 
the present above scenario is considered to be also kept in multi-band superconductors. 
The realistic calculation for iron-based superconductors is a future work.

\section{Conclusion}

In conclusion, we self-consistently performed large-scale BdG calculations in the presence of
vortices together with random impurities by using the Chebyshev-polynomial expansion
scheme to investigate QPI in the magnetic field.  
Our calculations concluded that the a vortex core distorted by impurity works as an  
impurity breaking the time reversal symmetry.
The microscopic finding well explains 
the observed experimental data, e.g., the QPI peaks relevant to the sign-preserving quasi-particle 
scattering grows while the sign-reversing one diminishes with increasing the magnetic field. 
Our direct calculations confirmed that the magnetic field dependence of QPI is 
a true phase-sensitive tool for unconventional superconductors.

%%%%%%%%%%%%%%%%%%%%%%% 3/16 10:43 %%%%%%%%%%%%%%%%%%%%%%%%%%%%%%

We would like to thank Y. Ota and  R. Igarashi for helpful discussions.  
The calculations have been performed using the supercomputing 
system PRIMERGY BX900 in Japan Atomic Energy Agency. 

%%%%%%%% 7/11 1:42 %%%%%%%%%%%%%%%%%%%%%%%%%%%%%%%%%%%%

%\pagebreak

\section*{references}


\begin{thebibliography}{99}
%\bibliography{apssamp}% Produces the bibliography via BibTeX.
\bibitem{Tsuei}
C. C. Tsuei and J. R. Kirtley, Rev. Mod. Phys. {\bf 72}, 969 (2000).
\bibitem{HanaguriFe}
T. Hanaguri, S. Nittaka, K. Kuroki, and H. Takagi, Science {\bf 328}, 474 (2010). 
\bibitem{Hoffman}
J. E. Hoffman, K. McElroy, D.-H. Lee, K. M. Lang, H. Eisaki, S. Uchida, and J. C. Davis, Science {\bf 297} 1148 (2002).
\bibitem{McElroy}
K. McElroy, R. W. Simmonds, J. E. Hoffman, D.-H. Lee, J. Orenstein, H. Eisaki, S. Uchida, and J. C. Davis, 
Nature (London) {\bf 422}, 592 (2003). 
\bibitem{Hanaguri}
T. Hanaguri, Y. Kohsaka, J. C. Davis, C. Lupien, I. Yamada, M. Azuma, M. Takano, K. Ohishi, M. Ono, and H. Takagi, 
Nat. Phys. {\bf 3}, 865 (2007). 
\bibitem{Kohsaka}
Y. Kohsaka, C. Taylor, P. Wahl, A. Schmidt, J. Lee, K. Fujita, J. W. Alldredge, K. McElroy, J. Lee, H. Eisaki, S. Uchida, D.-H. Lee, and J. C. Davis, Nature (London) {\bf 454}, 1072 (2008). 
\bibitem{HanaguriSci}
T. Hanaguri, Y. Kohsaka, M. Ono, M. Maltseva, P. Coleman, I. Yamada, M. Azuma, M. Takano, K. Ohishi, and H. Takagi, Science {\bf 323}, 923 (2009). 
\bibitem{Pereg-Barnea08}
T.~Pereg-Barnea and M.~Franz, Phys. Rev. B {\bf 78}, 020509(R) (2008).
\bibitem{Maltseva}
M. Maltseva and P. Coleman, 
%arXiv:0903.2752 (unpublished). 
Phys. Rev. B {\bf 80}, 144514 (2009). 
\bibitem{Nunner}
Tamara S. Nunner, Wei Chen, Brian M. Andersen, Ashot Melikyan, and P. J. Hirschfeld, Phys. Rev. B {\bf 73}, 104511 (2006). 
\bibitem{Zhang}Y. Y. Zhang, C. Fang, X. Zhou, K. Seo, W.-F. Tsai, B. A. Bernevig and J. Hu, 
%arXiv:0903.1694 (unpublished). 
Phys. Rev. B {\bf 80}, 094528 (2009). 
\bibitem{NagaiKato}
Y. Nagai and Y. Kato, Phys. Rev. B {\bf 82}, 174507 (2010). 
\bibitem{Balatsky}
A. V. Balatsky, I. Vekhter, J.-X. Zhu, Rev. Mod. Phys. {\bf 78}, 373(2006)
\bibitem{Wang}
Y. Wang, A. H. MacDonald, Phys. Rev. B {\bf 52} (1995) 3876(R). 
\bibitem{Covaci}
L. Covaci, F. M. Peeters, and M. Berciu, Phys. Rev. Lett. {\bf 105}, 167006 (2010).
\bibitem{Kunishima}
W. Kunishima, M. Itoh, and H. Tanaka, Prog. Theo. Phys. Supplement {\bf 138}, 149 (2000). 
\bibitem{Tanaka}
H. Tanaka, W. Kunishima, M. Itoh, RIKEN Review {\bf 29}, 20 (2000). 
\bibitem{NagaiOta}
Y. Nagai, Y. Ota, and M. Machida, 
J. Phys. Soc. Jpn. {\bf 81} (2012) 024710. 
% arXiv:1105.4939 (unpublished). 
\bibitem{Takigawa}
M. Takigawa, M. Ichioka, and K. Machida, J. Phys. Soc. Jpn. {\bf 69}, 3943 (2000). 
\bibitem{HanaguriPri}
T. Hanaguri, private communication.
\bibitem{note}
Such distortion is too subtle to observe in real-space DOS's, while observable
as spots in the Fourier QPI map. 






\end{thebibliography}
\end{document}